\documentclass[reprint,aps,prd,floats,floatfix,amsmath,amssymb,amsfonts,nofootinbib,longbibliography,superscriptaddress]{revtex4-2}

\usepackage{graphicx}% Include figure files
\usepackage{dcolumn}% Align table columns on decimal point
\usepackage{bm}% bold math
\usepackage{natbib}

\usepackage[hyperindex,colorlinks]{hyperref}

\renewcommand{\vec}{\boldsymbol}

\newcommand*{\mathabxbfamily}{\fontencoding{U}\fontfamily{mathb}\selectfont}
\DeclareFontFamily{U}{mathb}{\hyphenchar\font45}
\DeclareFontShape{U}{mathb}{m}{n}{
      <5> <6> <7> <8> <9> <10> gen * mathb
      <10.95> mathb10 <12> <14.4> <17.28> <20.74> <24.88> mathb12
      }{}
\newcommand*{\Earth}{{\text{\mathabxbfamily\char"43}}}
\newcommand*{\Moon}{{\text{\mathabxbfamily\char"4B}}}

\begin{document}

\title{Celestial mechanics in Newtonian-like gravity with variable $G$}

\author{Felipe S. Escórcio}
\affiliation{%
Departamento de F\'isica, Universidade Federal de Ouro Preto (UFOP), Campus Universit\'ario Morro do Cruzeiro, 35.400-000, Ouro Preto, Brazil}%

\author{J\'ulio C. Fabris}
\email{julio.fabris@cosmo-ufes.org}%
\affiliation{%
N\'ucleo Cosmo-ufes \& Departamento de F\'isica,  Universidade Federal do Esp\'irito Santo (UFES)\\
Av. Fernando Ferrari, 540, CEP 29.075-910, Vit\'oria, ES, Brazil.}%

\affiliation{National Research Nuclear University MEPhI (Moscow Engineering Physics Institute),
115409, Kashirskoe shosse 31, Moscow, Russia}

\author{Júnior D. Toniato}%
\email{junior.toniato@ufes.br}
\affiliation{%
N\'ucleo Cosmo-ufes \& Departamento de Química e Física, Universidade Federal do Espírito Santo - Campus Alegre, ES, 29500-000, Brazil}%

\author{Hermano Velten}%
\email{hermano.velten@ufop.edu.br}
\affiliation{%
Departamento de F\'isica, Universidade Federal de Ouro Preto (UFOP), Campus Universit\'ario Morro do Cruzeiro, 35.400-000, Ouro Preto, Brazil}%
\date{\today}

\begin{abstract}

A Newtonian-like theory inspired by the Brans-Dicke gravitational Lagrangian has been recently proposed in Ref. \cite{jjtv}. This work demonstrates that the modified gravitational force acting on a test particle is analogous to that derived from the Manev potential. Specifically, an additional term $\propto r^{-3}$ emerges alongside the conventional Newtonian component. We analyse the predicted expression for the pericenter advance and the Roche limit and use them to constraint the theory's single free parameter $\omega$ which is analogous to the Brans-Dicke parameter. At the same time this theory is able to solve the advance of Mercury's perihelion, we also show that there is no relevant impact on the Roche limit in comparison to the well known Newtonian results.
\end{abstract}

\maketitle

\section{Introduction}

The question of whether the fundamental constants of physics remain constant has been a long-standing inquiry. Four fundamental constants are particularly significant: $h$, defining the quantum world; $c$, the speed of light associated with relativistic effects; $G$, indicating gravitational interaction; and $k_B$, the Boltzmann constant related to thermodynamics. Of the four constants, $G$, was the first to be identified but is known with the least precision, only up to the order of $10^{-4}$ \cite{10.1093/nsr/nwaa165}. 

Modern theories interpret the gravitational phenomena as a consequence of spacetime curvature. Due to its dominant influence on large-scale systems like astrophysics and cosmology, constraints on the variation of $G$ from observations and experiments are extremely strict. Even a slight variation in $G$ with time or position can significantly impact cosmological and astrophysical observations, including addressing the Hubble constant tension \cite{PhysRevD.104.L021303} and altering the cosmological large scale structure formation scenario \cite{Alestas:2022xxm}. 

Various relativistic theories of gravity attempt to incorporate the variation of gravitational coupling, with the Brans-Dicke theory being a traditional paradigm \cite{Brans:1961sx}. The latter is a piece of the Horndesky class of theories that provides the most comprehensive gravitational Lagrangian, yielding to second-order field equations that often allow for a dynamical gravitational coupling \cite{Horndeski:1974wa}. These are examples of scalar-tensor theories where the gravitational interaction is mediated both by the metric tensor $g_{\mu\nu}$ as well as by a new dynamical scalar field $\phi\equiv \phi(t)$ which couples to the gravitational coupling $G$ i.e., $G\equiv G(\phi)$. This is an example of the physical mechanism behind this possible variation of $G$.

It is important to highlight that General Relativity (GR) stands as the dominant and widely accepted theory for describing the gravitational interaction. It has withstood rigorous scrutiny and demonstrated remarkable accuracy in explaining a myriad of gravitational phenomena. Over the years, it has successfully passed numerous tests at both the solar and Galactic levels, reaffirming its validity and precision in predicting the motion of planets, the bending of light around massive objects, and other gravitational effects.

However, despite its immense success, alternative theories to GR have been studied, hoping to refine or expand our understanding of gravity. One notable alternative is the Brans-Dicke theory, proposed as a way to incorporate a dynamical gravitational coupling. This theory offers deviations from GR for certain values of a dimensionless new coupling parameter, denoted as $\omega$. Notably, in the limit of extremely large $\omega$ values, the Brans-Dicke theory essentially reduces to the familiar framework of General Relativity.

Nonetheless, dark matter and dark energy dominate the behavior of the universe at the cosmological scale, yet their true nature remains elusive. Existing theories, including General Relativity, have struggled to account for their existence and effects adequately. As a result, the absence of a convincing explanation for dark matter and dark energy serves as motivation to probe the gravitational sector for signs of new physics.

In light of the so called dark sector a new investigation route has been established: to seek novel approaches and alternative theoretical frameworks for gravity. Exploring beyond General Relativity could potentially lead to breakthroughs, shedding light on the elusive nature of dark matter and dark energy. Such efforts might unveil new, yet undiscovered gravitational phenomena. While General Relativity has proven its worth and remains the leading theory for describing gravitational interactions, it is essential to explore alternatives like the Brans-Dicke theory to deepen our comprehension of gravity and its effects. Therefore, it is valid to seek for non-covariant based theories of gravity that can be directly compared to successfull Newtonian predictions.

Despite the abundance of relativistic theories with varying gravitational coupling, constructing a Newtonian theory with a dynamic $G$ is challenging. Initial proposals to incorporate a varying $G$ effect in a Newtonian context were relatively simple, replacing the constant $G$ in the Poisson equation with a varying gravitational coupling function $G(t)$ \cite{Duval:1990hj}. However, there is no dynamic equation governing this new function, necessitating ad hoc imposition of its behaviour with time. A natural choice is to employ the Dirac proposal \cite{Dirac:1937ti, Dirac:1938mt}, with $G = G_0(t_0/t)$, where $t_0$ is the present age of the universe and $G_0$ is the current value of the gravitational coupling. Nonetheless, this Newtonian theory with varying gravitational coupling lacks a complete Lagrangian formulation since $G(t)$ remains an arbitrary function.

In a recent paper \cite{jjtv}, a new Newtonian theory with a varying gravitational coupling has been proposed. The gravitational coupling, expressed as a function of time and position, is dynamically determined alongside the gravitational potential through a novel gravitational Lagrangian. This theory demonstrates consistency with the general properties of spherical objects like stars and can generate homogeneous and isotropic cosmological solutions leading to an accelerated expansion of the universe (see also \cite{Fabris:2021qkp}).

The interest in constructing a Newtonian theory with varying $G$ arises from several motivations. First, it satisfies the academic curiosity of developing a complete and coherent Newtonian formulation that incorporates a dynamic gravitational coupling. The Newtonian framework is conceptually simpler than the relativistic one, raising questions as to why providing $G$ with a dynamic behaviour is challenging, despite being feasible in a relativistic context. Second, many astrophysical and cosmological problems are more conveniently analysed within a Newtonian framework, including the dynamics of galaxies, galaxy clusters, and numerical simulations of large-scale structures. Having a consistent Newtonian theory that accounts for a non-constant $G$ would be desirable in such cases.

Our main goal in this work is to obtain in a post-Newtonian inspired way the effective gravitational force acting of a system of massive bodies or individual test particles. This has not been done previously in the sequence of works by some of the same authors. 

In this work we apply the non-relativistic prototype of varying $G$ theory developed in \cite{jjtv} to the celestial dynamics domain by focusing on the orbital pericenter advance,\footnote{Pericenter is a generic term used to refer to the point of maximum approximation between two celestial bodies in a bounded orbit. When considering orbital motion of planets around the Sun, this point is also called perihelion.} with particular application to the Mercury's Perihelion precession, but also the modified Roche limit expression for this theory. Our aim is to verify whether or not one can constrain the free parameter $\omega$ from this local analysis.   

The general theoretical set up is reviewed in the beginning of the next section. We then proceed calculating the corrections in the gravitational potential expansion up to second order. The equations of motion and applications to the pericenter advance are performed in section III.  We discuss the equivalence principle in section IV. In sections V and VI the Roche limit is analysed. We conclude in the final section.

\section{The gravitational force in varying $G$ Newtonian gravity}\label{sec:force}
The Newtonian theory with a variable $G$ is obtained through the following Lagrangian \cite{jjtv},
\begin{equation}\label{LagrangianG}
{\cal L} = -\frac{\nabla\psi\cdot\nabla\psi}{8\pi G_0} + \frac{\omega}{8\pi G_0}\biggr(\psi\frac{\dot\sigma^2}{\sigma^2} - c^4\nabla\sigma\cdot\nabla\sigma \biggl) - \,\rho \sigma \psi,
\end{equation}
where $G_0$ is a constant, $\psi$ is an equivalent of the ordinary Newtonian potential and $\sigma$ is a new dimensionless dynamical field. The Newtonian limit is fully recovered by setting $\sigma=1$. The field $\sigma$ is supposed to play the same rôle as the scalar field in covariant scalar-tensor gravitational theories. Here, the parameter $\omega$ is also a constant, the theory's free parameter. While not obligatory, it provides utility in facilitating the final dynamics and can also be interpreted as the analogous Brans-Dicke parameter. Concerning $\omega$, however, as we will see, the Newtonian limit is achieved with $\omega\rightarrow \infty$. A constant with dimensions of velocity, the speed of light $c$, has been introduced to guarantee the Lagrangian has the correct physical dimensions. However, in our reasoning we are not making direct mention to a relativistic framework in doing so: actually, this is equivalent to borrow from electromagnetism the two fundamental constants, the vacuum electric permitivity $\epsilon_0$ and the magnetic permeability $\mu_0$.

The field equations of the theory are obtained by varying the Lagrangian with respect to the fields $\psi$ and $\sigma$. They read
\begin{align}
\label{en1}
\nabla^2\psi + \frac{\omega}{2}\left(\frac{\dot\sigma}{\sigma}\right)^2  &= 4\pi G_0 \sigma \rho,\\[1ex]
\label{en2}
\nabla^2\sigma - \frac{1}{c^4\sigma}\frac{d}{dt}\left(\frac{\psi\dot\sigma}{\sigma}\right) &= \frac{4\pi G_0\psi \rho}{c^4\omega}.
\end{align}
The over-dot indicates total time derivative, which assures to the resulting equations invariance with respect to Galilean transformations. Now, from the above equations one can see that the Newtonian limit of the theory i.e., the Poisson equation is obtained both when $\sigma$ is a constant and $\omega\rightarrow\infty$ simultaneously.

In order to examine the dynamics of a massive body we verify from \eqref{LagrangianG} that the interaction of the gravitational fields with matter is given by the term $\rho\psi\sigma$. This will give rise to an effective gravitational potential such that Euler's equation for a self-gravitating fluid becomes,
\begin{equation}\label{euler}
    \rho\frac{d\vec{v}}{dt}=-\rho \vec\nabla(\psi\sigma)-\vec\nabla p.
\end{equation}
Thus, one can define the effective gravitational potential in the varying $G$ Newtonian gravity as
\begin{equation}\label{u-eff}
    U_{\rm eff} = -\psi\sigma+ \lambda,
\end{equation}
with $\lambda$ an arbitrary constant which does not contribute to the equation of motion of massive bodies.

\subsection{Corrections to the Newtonian potential}
In this section we will analyse the structure of the $U_{\rm eff}$ potential using an approximate method in order to obtain the first order corrections to the ordinary Newtonian gravitational potential. Let us expand the scalar fields up to second order,
\begin{align}
    \psi &=\psi_0 +\psi_1 + \psi_2 + ... \\
    \sigma &=\sigma_0 +\sigma_1 + \sigma_2 + ... \,,
\end{align}
where $\psi_0$ and $\sigma_0$ are constants and $\psi_2\ll\psi_1\ll\psi_0$ as well as $\sigma_2\ll\sigma_1\ll\sigma_0$. Moreover, in substituting this into the field equations we will also assume that every time derivative is one order higher, i.e. $\dot{\psi}_1\sim O(2)$ (this is equivalent to assume that the gravitational fields are slowly varying in time, which is consistent with local environments as the solar system), and the matter density $\rho$ is $\sim O(1)$ (since it is responsible to create the perturbed fields).

\subsubsection{First order}
Retaining only first order terms, one obtains
\begin{align}
    \nabla^2\psi_1&=4\pi G_0\sigma_0\rho,\\[1ex]
    \nabla^2\sigma_1&=\frac{4\pi G_0}{c^4\omega}\,\psi_0\rho.
\end{align}
The solutions of these equations are proportional to the ordinary Newtonian potential $U$,
\begin{align}
    \psi_1=-\frac{G_0\sigma_0}{G_N} \,U,\quad \sigma_1=-\frac{G_0 \psi_0}{G_Nc^4\omega} \,U,
\end{align}
with
\begin{equation}\label{newt}
    U=G_N\int\frac{\rho(x')}{\vert \vec{x}-\vec{x}{}'\vert}\,d^3x'.
\end{equation}
being the negative of the Newtonian potential and $G_N$ is the Newtonian gravitational constant.

The effective gravitational potential then becomes,
\begin{equation}\label{Ueff}
    U_{\rm eff}=-\psi\sigma +\lambda\approx \frac{G_0}{G_N}\left( \frac{\psi_0^2}{c^4\omega}+\sigma_0^2\right)U,
\end{equation}
where $\lambda$ has been set equals to zero without loss of generality. Since it is expected that the second scalar field $\sigma$ introduces only corrections to the Newtonian gravity, one can also consider the approximation
\begin{equation}
    \frac{\psi_0^2}{c^4\omega}\ll\sigma_0^2.
\end{equation}
Thus, after taking this limit in \eqref{Ueff}, at first order, the theory will reproduce Newtonian gravity if
\begin{equation}\label{G0}
    G_0\sigma^2_0=G_N.
\end{equation}
For simplicity, let us work with $\sigma_0=1$ (there is no new physics in doing this choice, because any value of this constant can be always absorbed by the coupling constant $G_0$.

\subsubsection{Second order}
Using the previous results, the second order field equations are
\begin{align}
    \nabla^2\psi_2 &=-\frac{4\pi G_N\psi_0}{c^4\omega}\,\rho U,\\[1ex]
    \nabla^2\sigma_2 &=-\frac{4\pi G_N}{c^4\omega}\,\rho U,.
\end{align}
Using the post-Newtonian potential,
\begin{equation}
    \Phi_2=G_N\int\frac{\rho(x')U(x')}{\vert \vec{x}-\vec{x}{}'\vert}\,d^3x',
\end{equation}
one has
\begin{align}
    \psi_2 &=\frac{\psi_0}{c^4\omega}\,\Phi_2\\[1ex]
    \sigma_2 &=\frac{1}{c^4\omega}\,\Phi_2,
\end{align}
after using definition \eqref{G0}.

Finally, going to the second order of the effective potential, one obtains
\begin{equation}
    U_{\rm eff}\approx U - \frac{\psi_0}{c^4\omega}(U^2+2\Phi_2).
\end{equation}
It is worth noting that both second order potentials, $U^2$ and $\Phi_2$, are present in the post-Newtonian metric of general relativity \cite{poisson_will_2014}.

\subsection{Equation of motion of massive bodies}
We are interested in deriving the equation of motion of a certain body $A$ within a N-body system. The force acting on this specific body reads
\begin{equation}
\vec{f}_A=m_A\vec{a}_A=\int_A\rho(t,\vec{x})\,\frac{d\vec{v}}{dt}\,d^3x.    
\end{equation}
In the above expression, the domain of integration is a fixed volume $V_A$, which is bigger than the volume occupied by body $A$ but also smaller enough such that it does not intersects any other body of the system. This is feasible once we will assume the bodies are far away apart from each other. Moreover, $\vec{a}_A$ stands for the center-of-mass acceleration of body $A$.

Plugging Euler's equation \eqref{euler} into the integral above, one obtains 
\begin{equation}\label{int}
\vec{f}_A= \int_A\rho(t,\vec{x})\vec{\nabla}U_{\rm eff}d^3x.
\end{equation}
The pressure contribution is found to be zero by using the Gauss theorem and noting that $p=0$ in the border of the volume of integration. Once the gravitational potential within the integral is due to the contribution of all bodies within the system, including body $A$, we need to split it in two parts. For instance, we rewrite the Newtonian potential as
\begin{equation}
    U=U_A + U_{{\rm ext},A},
\end{equation}
where $U_A$ stands for the potential created by the body $A$,
% \begin{equation}
%     U_A=G_N\int_A\frac{\rho(t,\vec{x}')}{\x}\,d^3x'
% \end{equation}
and $U_{{\rm ext,A}}$ is the contribution due to the remaining bodies of the system.
% \begin{equation}
%     U_{{\rm ext,A}}=G_N\sum_{B\neq A}\int_A\frac{\rho(t,\vec{x}')}{\x}\,d^3x'.
% \end{equation}
Both are calculated as in \eqref{newt}, but $U_A$ is integrated over $V_A$ while the external potential is a sum of integrals over the volume surrounding each one of the remaining bodies of the system. By considering that each body density is reflect symmetric over its center of mass, it is possible to show that only the external parts do contribute to the force. Moreover, these terms can be removed from the integral in \eqref{int} since the bodies are wide separated, i. e., the external potentials will not depend on the integration variable as a first approximation.\footnote{The results for the integral of the divergence of each potential is already known. Reference \cite{poisson_will_2014} presents a well detailed exposition of those calculations. Although it is used a distinct definition of mass density, since the conserved density in general relativity is given by $\rho*=\rho\sqrt{-g}$, their results are still valid in the Newtonian context once here $\rho$ is the one satisfying the continuity equation.} After the above considerations, one has
\begin{align}
    f_A^j= \ m_A&\partial_jU_{{\rm ext},A} -\frac{2\psi_0m_A}{c^4\omega}\,\partial_j\Phi_{2\,{\rm ext},A} \ + \notag\\[1ex]
    &\,\frac{2\psi_0\partial_jU_{{\rm ext},A}}{c^4{\omega}}\left(2\Omega_A -m_AU_{{\rm ext},A} \right),\label{force}
\end{align}
where $\partial_j\equiv \partial/\partial x^j$, $m_A$ is the mass of body $A$ and $\Omega_A$ is its gravitational energy, namely
\begin{align}
    m_A &=\int_A\rho \,d^3x,\\[1ex]
    \Omega_A &= -\frac{G_N}{2}\int_A\frac{\rho(\overline{x})\rho(\overline{x}') }{\vert \vec{\overline{x}}-\vec{\overline{x}'}\vert}\,d^3\overline{x}d^3\overline{x}',
\end{align}
with $\vec{\overline x}=\vec x - \vec{r}_A$ being the body coordinate with respect to its center-of-mass $\vec r_A$. 
%The indices $j$ and $k$ run from 1 to 3, indicating spatial components. Thus, using Einstein notation, $\Omega_A=\Omega_A^{jk}\delta_{jk}$ is the body gravitational energy.

Once we are assuming the bodies are well separated, the external potentials can be expanded in a Taylor series (see, for instance, \cite[pp. 437]{poisson_will_2014}),
\begin{align}
    U_{{\rm ext},A} &= \sum_{B\neq A}\frac{G_Nm_B}{r_{AB}},\\[1ex]
    \partial_jU_{{\rm ext},A} &= -\sum_{B\neq A}\frac{G_Nm_B}{r_{AB}^2}\,\hat{r}_{AB}^j,\\[1ex]
    \partial_j\Phi_{{\rm ext},A} &= 2\sum_{B\neq A}\frac{G_N\Omega_B}{r_{AB}^2}\,\hat{r}_{AB}^j - \sum_{B\neq A}\frac{G_N^2m_Am_B}{r_{AB}^3}\,\hat{r}_{AB}^j - \notag\\[1ex]
    & \quad \sum_{B\neq A}\sum_{C\neq A,B}\frac{G_N^2m_Bm_C}{r_{AB}^2r_{BC}}\,\hat{r}_{AB}^j.
\end{align}
In the above expression we are using the notation $\vec{r}_{AB}=\vec{r}_A - \vec{r}_B$ and $\hat{r}_{AB}=\vec{r}_{AB}/r_{AB}$.

With a redefinition of the mass of each body,
\begin{equation}\label{mass}
    M_A=m_A + \frac{4\Omega_A}{c^2\,{\widetilde\omega}}
    %=m_A\left(1+\frac{4 \Omega_A}{m_A c^2 \widetilde{\omega}}\right),
\end{equation}
with
\begin{equation}\label{widetilde}
    \widetilde{\omega}\equiv c^2\omega/\psi_0,
\end{equation}
the final form of the force \eqref{force} will be
\begin{align}
    \vec{f}_A & = -\sum_{\scriptscriptstyle B\neq A}\frac{G_NM_AM_B}{r_{AB}^2}\,\vec{\hat{r}_{AB}} \ + \notag\\[1ex]
    &  \sum_{\scriptscriptstyle B\neq A}\frac{G_N^2M_AM_B}{c^2{\,\widetilde{\omega}}\,r_{AB}^2}\frac{(M_A+M_B)}{r_{AB}}\,\vec{\hat{r}_{AB}} \ + \notag\\[1ex]
    &   \sum_{\scriptscriptstyle B\neq A}\sum_{\scriptscriptstyle C\neq A,B}\!\!\frac{G_N^2M_AM_BM_C}{c^2{\,\widetilde{\omega}}\,r_{AB}^2}\left(\frac{1}{r_{AC}}+\frac{1}{r_{BC}}\right)\vec{\hat{r}_{AB}},\label{force-final}
\end{align}
since terms of order $\psi_0^2/c^4$, or higher, are neglected.

%{\bf The term $4\Omega_A/m_A c^2 \widetilde{\omega}$ can be interpreted as the deviation from the equivalence principle in this theory. Given the recent results obtained by the MICROSCOPE mission this deviation is limited to the order $10^{-15}$ \cite{microscope}. Therefore, since the ratio $\Omega_A/m_A$ for systems like Earth and Moon is $\sim 10^{-10}$ and assuming the $c^2\sim 10^{17}$ one can estimate $\widetilde \gtrsim 10^{-12}$ as the lower bound for this quantity.} ESSA ANÁLISE FOI ERRADA PORQUE, NA VERDADE, É $\Omega_A/m_Ac^2\sim 10^{-10}$. OU SEJA, FALTOU O $c^2$, O QUE MUDA TUDO.

%Since $\psi_0$ is the zeroth order gravitational potential solution its value in a bounded system is negative. Therefore, $\tilde{\omega}$ is expected to assume negative values. The Newtonian limit will be recovered with $\tilde{\omega}\rightarrow - \infty$.

The first term in \eqref{force-final} resembles the typical Newtonian gravitational force in a $N$-body system. The distance dependence is the ordinary one but the departure from Newtonian gravity comes from the mass redefinition \eqref{mass}. It shows a contribution of the body gravitational energy in the gravitational attraction. This is a remarkable difference between this formalism and Newtonian gravity.  It indicates a distinction between inertial and gravitational masses in this theory. The first one is $m$, the mass relating force and acceleration, while the latter, $M$, is the mass determining the gravitational potential (in relativistic terms, the active gravitational mass). But it must be noted that this departure does not violate the Newtonian limit of the theory, since it only occurs at the post-Newtonian order.

It is worth to note also that in GR, a mass redefinition in post-Newtonian analysis is present too. But there $(M_A)_{\scriptscriptstyle\rm GR}=m_A+E_A/c^2$, where $E_A$ is the body total energy, and this redefinition does not spoil the mass conservation statement. Neither the equivalence between inertial and gravitational mass is broken. Notwithstanding, some others relativistic theories of gravity can eventually violate this equivalence principle, even after the relativistic mass redefinition. Even so, the distinction between inertial and gravitational mass is well constrained now a days and this must give a upper bound to the parameter $\widetilde\omega$. In Section \ref{sec:ep} we will derive this constrain.

\subsection{Binary systems}
Considering a system of two bodies, the force \eqref{force-final} can be simplified by using the relative motion between the bodies with respect to the system center-of-mass. Therefore, the binary system can be interpreted as a single particle of reduced mass $\mu=M_1M_2/M$ moving in the gravitational field of a body of mass $M=M_1+M_2$. However, one can not measure the difference in gravitational and inertial mass in a two-body system. Thus, we can neglect the distinction between $M$ and $m$ and assume the latter as the Kepler-measured mass of the system. 

With that said, the force in a two-body system, up to first post-Newtonian order, can be written as follows,
\begin{equation}
\vec{f}=\left(-\frac{G_Nm\mu}{r^2} + \frac{G_N^2m^2\mu}{c^2{\,\widetilde{\omega}}\,r^3}\right)\vec{\hat{r}},\label{force-pert}
\end{equation}
with $\mu=m_1 m_2 / m$ and $m=m_1+m_2$ and also
$\vec r=\vec r_1 - \vec r_2$. Therefore, it is noteworthy
that the smaller the free constant $\widetilde{\omega}$ is, the larger the perturbation term of the force due to variable G will be.

\section{Variations of orbital elements}
\begin{figure}[t!]
    \centering
    \includegraphics[width=\linewidth]{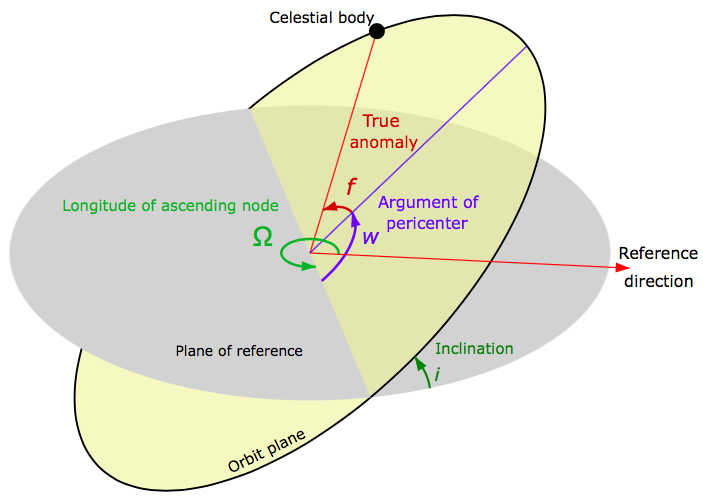}
    \caption{Orbital motion decomposed in a reference plane. Credits: \href{https://en.wikipedia.org/wiki/User:Lasunncty}{Lasunncty}, through \href{https://commons.wikimedia.org/wiki/File:Orbit1.svg}{Wikimedia commons}. Used under license \href{https://creativecommons.org/licenses/by-sa/3.0/deed.en}{CC BY-SA 3.0} and adapted by the authors.}
    \label{fig:orbit}
\end{figure}

The modified gravitational force obtained in the previous section will not make test particles to follow perfect ellipses around the source. Actually, the Keplerian parameters which characterise an orbit and its orientation with respect to the central object will undergo through secular variation. The most iconic of those variations is the advance of the pericenter argument, the angle correspondent to the maximum approximation in a two-particle system. In order to obtain expressions for the variations of the Keplerian parameters, we interpret the correction to the Newtonian gravitational force as a perturbative effect. This is justified by the fact that, for bounded systems, the Virial theorem imply $G_Nm/r\sim v^2$, where $v$ is the typical orbital velocity. Thus, the correction to the Newtonian gravitational potential is at least of order of $v^2/c^2$, which is small for planetary or satellite orbits, for instance. We are also assuming that $\widetilde{\omega}$ is not too small to guarantee validity of the perturbative expansion.

The zeroth order solution will be the same as in Newtonian gravity. Considering a system of two particles with masses $m_1$ and $m_2$, and adopting the center-of-mass as the origin of the coordinate system, one thus obtain the conic equation for the trajectory of the relative position between the particles,
\begin{equation}\label{conic}
    r = \frac{a(1-e^2)}{1+e\,\cos{f}}.
\end{equation}
In the above, $e$ is the eccentricity, $a$ the semi-major axis and $f$ is the true anomaly, the orbital angle as measured from the pericenter. For observational purposes, it is necessary to describe the orbital motion with respect to the Earth's orbital plane, which requires the introduction of additional orbital parameters: the orbit inclination $\iota$, the angle $\Omega$ between the ascending node and a reference direction in the reference plane, and the argument of the pericenter $w$. These parameters are represented in Figure \ref{fig:orbit}.

To calculate the secular variations of the Keplerian parameters we use the method of osculating orbits. The particle is interpreted to being moving in a ellipse in each instant of time, but the orbital parameters defining the ellipse are time varying functions due to the perturbing acceleration. The rate of change of each orbital element is proportional to the acceleration components vector: an orbital radial component $\cal R$, a component $\cal W$ normal to the orbital plane, and a third component $\cal S$ orthogonal to the previous two. We refer to Reference \cite{poisson_will_2014} for more details on osculating orbits method.

The gravitational force \eqref{force-pert} resembles the Newtonian one plus an extra term proportional to $\widetilde{\omega}^{-1}$. If this extra term is interpreted as a perturbative correction, this leads to an acceleration correction with radial component only such as
\begin{equation}
    {\cal R} = \frac{1}{{\,\widetilde{\omega}}}\frac{G_N^2m^2}{c^2r^3}.
\end{equation}
This causes variations just on the eccentricity and pericenter argument, which are given by
\begin{align}
    \frac{dw}{df} =& \ - \frac{a^2(1-e^2)^2}{eGm}\frac{\cos f}{(1+e\cos f)^2}\,{\cal R}\\[1ex]
    \frac{de}{df} =& \  \frac{a^2(1-e^2)^2}{Gm}\frac{\sin f}{(1+e\cos f)^2}\,{\cal R}.
\end{align}
Using the relation \eqref{conic}, we integrate the above expressions over a complete orbital period to obtain
\begin{align}
    \Delta w =& \ -\frac{\pi}{{\,\widetilde{\omega}}}\frac{G_Nm}{a c^2(1-e^2)},\\[1ex]
    \Delta e =& \ 0.
\end{align}
Thus, from all the Keplerian orbital parameters, only the pericenter argument undergoes a secular variation. It is worth to note that general relativity prediction for the orbital variation of the pericenter advance is given by
\begin{equation}
    \Delta w_{\rm GR} = 6\pi\frac{G_Nm}{a c^2(1-e^2)}.
\end{equation}
This shows that the result derived from the modified Newtonian gravity with variable $G$ is the same as the general relativistic one if $\,\widetilde\omega=-1/6 \approx -0.16$.

The variation per orbit can be converted in a variation per time by dividing $\Delta w$ by an orbital period $P$. In this process, we can eliminate the period through Kepler's third law, i.e. $P=2\pi a^{3/2}/\sqrt{G_Nm}$.\footnote{In principle, Kepler's third law would be modified due to the new gravitational force but, since $\Delta w$ is already of post-Newtonian order, the corrections are negligible.} The result is
\begin{equation}\label{wsec}
    \left(\frac{dw}{dt} \right)_{\rm sec} = -\frac{1}{2\,\widetilde{\omega}}\frac{(G_Nm)^{3/2}}{a^{5/2}c^2(1-e^2)}.
\end{equation}

\subsection{Observational Constraints}\label{constraints}
The secular advance of the orbital pericenter can be used to obtain observational constraints for the parameter $\widetilde{\omega}$. In this section we will consider data from three systems where this secular variation is detected. In practise, we shall convert the existing observational results in constraints on the parameter $\widetilde{\omega}$.

\subsubsection{Perihelion precession of Mercury}
The most famous test of an advance of orbital pericenter is the one performed with Mercury. It consists of a classical gravitational test and it was determinant for the success of general relativity. The argument of Mercury's perihelion undergoes a secular variation due to the gravitational attraction of other planets from solar system, the non-zero quadrupole moment of the Sun and the precession of the Earth's spring equinox (the reference axis). However, all these contributions together can not explain a deficit given by,
\begin{equation}\label{wobs}
    \left(\frac{dw}{dt} \right)_{\rm sec} = (42.9799 \pm 0.0009)''/\mbox{century}.
\end{equation}
The above data was extracted from most recent estimations using MESSENGER spacecraft data \cite{Park_2017}. Others published values for this quantity can be consulted in Refs. \cite{Krasinsky:1993,Pitjeva:2001}.

For a numerical estimation of \eqref{wsec} we use $G_N = 6.674\times 10^{-11}$\,m$^3$/kg$\cdot$s$^2$, $M = M_{\odot} = 1.988\times 10^{30}\,$kg as the mass of the Sun, Mercury's eccentricity and semi-major axis $e = 0.205631$ and $a = 57.909\times10^9$\,m. Equality of \eqref{wsec} and \eqref{wobs} results in
\begin{equation}
    \,\widetilde\omega = -0.166591 \pm 0.000003.
\end{equation}

\subsubsection{LAGEOS Satellites}
The LAGEOS project consists of two satellites orbiting around the Earth with a precise tracking of their trajectories. LAGEOS II has higher eccentricity, which gives a more precise measurement of its pericenter advance. According to reference \cite{Lucchesi:2010}, the observed value is
\begin{equation}
    \left(\frac{dw}{dt} \right)_{\rm sec} = (3.351 \pm 0.007)''/\mbox{yr}.
\end{equation}
To evaluate the theoretical prediction we use $M = 5.9724\times10^{24}\,$kg for the mass of Earth, and LAGEOS II orbital parameters $e=0.0135$ and $a = 12.162\times10^6$\,m. The resulting bound is given by
\begin{equation}
    \,\widetilde\omega = -0.1667 \pm 0.0003.
\end{equation}

\subsubsection{Precession of S2 around Sgr A*}
After almost 30 years observing the motion of stars around the Sgr A$^*$, the black hole in the center of Milk Way galaxy, the GRAVITY collaboration has been able to determine the orbital precession of the star $S2$ to be \cite{Abuter:2020}
\begin{equation}
    \left(\frac{dw}{dt} \right)_{\rm sec} = (50 \pm 9)''/\mbox{yr}.
\end{equation}
The correspondent bound is given by
\begin{equation}
    \,\widetilde\omega =- 0.15 \pm 0.03,
\end{equation}
where we have taken $M = 8.26\times 10^{36}$\,kg, $e = 0.884649$ and $a = 1.54\times10^{14}$\,m.

\section{The equivalence principle}
\label{sec:ep}

Within ordinary Newtonian theory the relation between inertial and gravitational mass is given by the experiments: no theoretical relation exists between the two masses. They are considered equal because the experimental bounds suggests this equality.
General Relativity, on the other hand, make both masses equal from the beginning, constituting one of the cornerstones of the theory. For known modifications of the Newtonian theory the situation may acquire new features. McVittie theory behaves like Newtonian theory in this subject, while MOND \cite{Famaey2012,Milgrom2015,Bekenstein1984}, on the other hand, does not satisfy the equivalence principle in general (by the way, neither Newton's third law). 

The Newtonian theory with variable $G$ investigated here allows to have a theoretical prevision for the equivalence principle in terms of the free parameter $\widetilde\omega$. Of course, we may not expect this theory to satisfy the experimental bounds for the equality of inertial and gravitational mass, since it is not a relativistic theory. However,  we can obtain an estimation on the relation between both masses.

The dependence of a massive-body equation of motion with its internal structure, evidenced through the mass redefinition \eqref{mass}, can produce violations of the weak equivalence principle (WEP) at post-Newtonian order. This is better explained using a general quasi-Newtonian second law expression for a $N$-body system,
\begin{equation}\label{quasi-N}
    \vec{f}_A={(\cal M_I)}_A\vec a_A={(\cal M_P)}_A\vec{\nabla} \sum_{B\neq A}\frac{G_N{\cal (M_A)}_B}{r_{AB}}.
\end{equation}
In the above expression, ${\cal M_A}$ is the amount of mass generating a gravitational potential: the \textit{active gravitational mass}; ${\cal M_P}$ is the mass term relating force with the gradient of the potential: the \textit{passive gravitational mass}; and the \textit{inertial mass}, i.e. the relation between force and acceleration, is represented by ${\cal M_I}$.

An equality between inertial and passive gravitational masses is the manifestation of the WEP: the acceleration of a body in an external gravitational field does not depend of its mass or internal structure. This is the case happening here. Comparing \eqref{quasi-N} with the first term of \eqref{force-final}, one can easily make the identifications ${\cal M_I}=m$ and ${\cal M_P}={\cal M_A}=M$. The equivalence between active and passive gravitational masses is a consequence of the validity of Newton's third law, i.e., total momentum conservation of the system. 

%Violations of the WEP could have manifest in a system where two bodies are influenced by the gravitational field of a third body: the so called Nordtvedt effect. This is the case, for instance, of the Earth-Moon system while orbiting the Sun. If the Earth and Moon fall with different acceleration toward the Sun, their relative distance would change with time in a specific frequency that make possible to isolate this effect to other perturbations after a long period observation of the Moon's orbit \cite{poisson_will_2014}.

Tests of WEP are quantified by the Eötvos ratio,
\begin{equation}
    \eta=2\,\frac{{\cal M_P}-{\cal M_I}}{{\cal M_P}+{\cal M_I}}\approx \frac{4}{\widetilde\omega}\left(\frac{\Omega_2}{m_2c^2}-\frac{\Omega_1}{m_1c^2}\right),
\end{equation}
with the last terms being the result for the varying-G theory, and the sub-indexes 1 and 2 refereeing to generic bodies. The strongest constraint were obtained recently by the MICROSCOPE experiment, giving $\eta\lesssim 10^{-15}$ \cite{microscope}. The experiment is to measure the relative acceleration of two test masses, with different compositions, while both are orbiting the Earth. However, we couldn't find information on the values of $\Omega/m$ for the used test masses.

Notwithstanding, in Ref. \cite{Hofmann_2018}, improvements on lunar laser ranging analysis of the Earth-Moon system falling into the Sun gravitational potential determined the Eötvos ratio only about a factor of 10 weaker than MICROSCOPE result. With the difference of the gravitational self-energy between the Earth (\Earth) and Moon (\Moon) being
\begin{equation}
    \left(\frac{\Omega}{mc^2}\right)_{\Earth}\!\!-\left(\frac{\Omega}{mc^2}\right)_{\Moon} = -4.45\times 10^{-10},
\end{equation}
we can estimate, for $\widetilde\omega=-1/6$, that $\eta\sim 10^{-8}$. As expected, such violation of the experimental results is in the post-Newtonian order. It is worth to note that, to be in a complete agreement with those observational bounds, one must work with $\widetilde\omega\gtrsim 10^{5}$, which consequently would result in no corrections to the orbital pericenter shift.

\section{CORRECTIONS IN THE ROCHE LIMIT}
Modifications in gravitational force can alter the calculation performed for the Roche Limit, which establishes, through Newtonian mechanics, a critical distance for an object's orbit to not be sufficiently close to its host body, as this proximity would result in the satellite's disintegration. In this chapter, we are proceeding with the calculation of the modified Roche Limit, considering the contribution of the total gravitational force as presented in the last section. In the proposed calculation, the zeroth-order solution remains the same as in Newtonian gravity. However, we will demonstrate that the suggested correction (with $\widetilde\omega = -1/6$ for the advance of the pericenter) has minimal impact on the known Roche Limit, indicating that the developed theory addresses issues related to general relativity without affecting the already known numerical results in Newtonian physics.

To construct the Roche Limit affected by first-order force corrections, it is sufficient to consider that at the point of disintegration, the difference $\Delta g$ between the acceleration perceived by the edge of the satellite relative to its center will be given by equation \eqref{force-pert}. Thus, the difference in acceleration takes the form:
\begin{equation}
\Delta g = -\frac{2G_NM_hr_s}{r^3} + \frac{3G_N^2M_h^2r_s}{c^2 \widetilde\omega r^4}.
\end{equation}
When considering that at the brink of disintegration, this same difference will be approximately equal to the force that holds the satellite together, we obtain
\begin{equation}
-\frac{G_Nm_s}{r_s^3} + \frac{G_N^2m_s^2}{c^2\widetilde\omega r_s^4} = -\frac{2G_NM_h}{r^3} + \frac{3G_N^2M_h^2}{c^2\widetilde\omega r^4}. \label{eq}
\end{equation}
Therefore, in this case, we are stating that the distance $r$ between both bodies will be the upper limit for the difference in acceleration between their ends not to cause the satellite's disintegration. We will refer to this limit as the Variable-G Roche Limit. 

To implement equation \eqref{eq}, one can rewrite it as

\begin{equation}
-\frac{m_s}{r_s^3} + \frac{G_N}{c^2\widetilde\omega}\frac{m_s^2}{r_s^4} = -\frac{2M_h}{r^3} + \frac{G_N}{c^2\widetilde\omega}\frac{3M_h^2}{r^4}. \label{a1}
\end{equation}
This approach allows us to investigate how changing the scale of the constant $\widetilde{\omega}$ affects the behavior of the system and its physical properties. The aim here is to allow the comparison of different possible values for $\widetilde{\omega}$, without restricting it only to $-1/6$ and, thus, compare with the expected values for the Roche limit already known in the literature for some astrophysical systems. Let us investigate how the different $\widetilde{\omega}$ values impact the dynamics in different astrophysical scenarios. In this attempt, in the next section we will obtain the Roche limits as a function of  the $\widetilde{\omega}$ values fixing the real value of the analyzed satellite radius, which we refer to as $r_s$. Again, the typical Newtonian result is achieved in the limit $|\widetilde\omega| \rightarrow \infty$, now accompanied by a correction term proportional to $r^{-4}$. A perturbative term of this order can contribute to a slight alteration in the known result, but if the value assigned to $\widetilde{\omega}$ is sufficiently small, the correction can become significantly larger.

\section{Host-Satellite Systems}
The selection of the analyzed systems was based on evaluating their proximity to the Roche limit. With the exception of the celestial bodies Earth and Moon, the remaining selected systems are positioned considerably close to the disintegration limit. Our goal, therefore, is to quantify the impact of different values of $\widetilde{\omega}$ on the Roche limit.

 Given the results provided in the last section, we therefore will start verifying  values around $\widetilde\omega \approx - 0.16$. Thus, in order to perform an analysis close to the values of interest, it is necessary to go back to equation \eqref{a1} and consider that 

\begin{equation}
\frac{G_N}{c^2\widetilde\omega} = - 4,44667 \times 10^{-27}.
\end{equation}
For the above numerical estimation we have used $G_N = 6.674\times 10^{-11}$\,m$^3$/kg$\cdot$s$^2$ and $c = 3.0 \times 10^8 m/s $. A slight increase in the scale associated of the constant $G_N/c^2\widetilde\omega$ also leads to a considerable decrease in the constant $\widetilde\omega$. For instance, in the next figures, by adopting $\widetilde\omega \approx -0.016$, we obtain the green curve that shows a greater divergence from the expected values compared to $\widetilde\omega \approx -0.16$, represented by the red curve. When we adopt different values for the constant $\widetilde\omega$, we also obtain new values for the Roche limit. In our case, we will compare such divergences until $\widetilde\omega \approx -0.0016$, which will be represented by the blue line in each figure.

 \subsection{Earth-Moon System}
The numerical range chosen for the analysis of the Earth-Moon system has a center corresponding to the average radius of the satellite, which in this case is approximately $1737.4$km. Therefore, it is established as the midpoint between the values $1687$km and $1787$km, which were selected as the lower and upper limits, respectively. This choice allows us to create a curve based on equation \eqref{a1} that represents the $\widetilde\omega$ values within the same highlighted range.

To investigate the minimal influence of the perturbative force in the Roche limit, we consider the already known value for the Earth-Moon system and compare it with the resulting theoretical predictions when adopting different $\widetilde\omega$ values.

\begin{figure}[t]
\centering
\includegraphics[width=\linewidth]{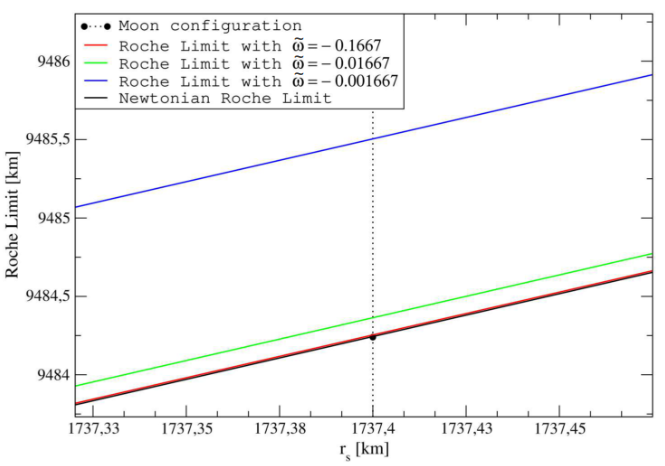}
\caption{Roche Limit Graph for different satellite radius values relative to a constant mass $m_s$. The different curves express the different values of $\widetilde\omega$. The dashed vertical line seeks to indicate the different Roche limits for $\widetilde\omega$ considering the radius of the Moon.}
\label{fig:moon}
\end{figure}

The Earth-Moon system consists of Earth, which has a mass of $5.97 \times 10^{-24}$ kg, and its natural satellite, the Moon, with a mass of $7.34 \times 10^{-22}$ kg. Upon analysing this case, it is observed that any value greater than $\widetilde\omega = - 0.1667$ leads to a significant departure from the expected results according to the classical Roche limit, represented by the black line (almost indistinguishable from the red line) in Fig. \eqref{fig:moon}. Nevertheless, choosing $\widetilde\omega = -0.1667$ provides a good approximation for the investigated system, considering that the theory with variable G should only cause a small disturbance to Newtonian gravity, remaining consistent with the values obtained in Section \ref{constraints}. In practise, values of order $|\widetilde\omega| \sim O( 10^{-1})$ have no impact in the Roche limit keeping the same value as in Newtonian gravity. Table \eqref{table1} refers to the numerical results obtained from this analysis. The presentation of our results in the next subsections follow the same structure.

\begin{table}[h]\label{table1}
    \centering
    \caption{Percentage differences in the Roche limit for different values of $\widetilde{\omega}$. This results quantifies the curves shown in Fig. \ref{fig:moon} for the Earth-Moon system. }
    \label{tab:moon}
    \begin{tabular}{c|c}
        $\widetilde\omega$ & $\text{Deviation from the Newtonian Roche limit}$ \\
        \hline
            $- 0.1667$ & $\approx 0.00011 \%$\\
            
            $- 0.01667$ & $\approx 0.0013  \%$\\
            
            $- 0.001667$ & $\approx 0.013\%$\\
        \end{tabular}
\end{table}

\subsection{Jupiter-Europa System}
The method employed to analyse the Jupiter-Europa system is the same as before, the difference lies in the values of the intrinsic orbital parameters of the system, where the average radius of the satellite in this case will be approximately $1560.8$km. Therefore, it is established as the midpoint between the values $1510$km and $1610$km, which were selected as the lower and upper bounds, respectively.

To investigate the minimal influence of the perturbative force in the Roche limit, we consider the already known value for the Jupiter-Europa system and compare it with the expected theoretical values when adopting different $\widetilde\omega$.

\begin{table}[htb]\label{table2}
\caption{Percentage differences in the Roche limit for different values of $\widetilde{\omega}$. This results quantifies the curves shown in Fig. \ref{fig:europa} for the Jupiter-Europa system}
    \label{tab:europa}
    \centering
    \begin{tabular}{c|c}
        $\widetilde\omega$ & $\text{Deviation from the Newtonian Roche limit}$ \\
        \hline
            $- 0.1667$ & $\approx 0.0063 \%$\\
            
            $- 0.01667$ & $\approx 0.063 \%$\\
            
            $- 0.001667$ & $\approx 0.62 \%$\\
        \end{tabular}
\end{table}
    
\begin{figure}[htb]
\centering
\includegraphics[width=\linewidth]{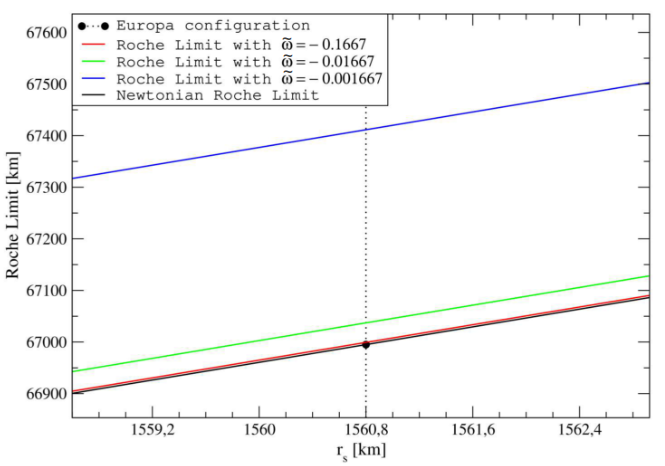}
\caption{Roche Limit Graph for different satellite radius values relative to a constant mass $m_s$. The different curves express the different values of $\widetilde\omega$. The dashed vertical line seeks to indicate the different Roche limits for $\widetilde\omega$ considering the radius of the Europa.}
\label{fig:europa}
\end{figure}

The Jupiter-Europa system consists of Jupiter, which has a mass of $1.898 \times 10^{27}$ kg, and its natural satellite, Europa, with a mass of $4.8 \times 10^{22}$ kg. Obviously, the constant $\omega$ does not directly depend on the orbital parameters in its equation, so any value smaller than $\widetilde\omega \approx -0.1667$ would again result in a divergence from the expected results according to the classical Roche limit.

It is observed that the orbital values directly influence the expected distance between the Roche limit with variable $G$ and the Newtonian Roche limit since the dependence on the parameters is inherent in equation \eqref{eq}. Even with the value of order $\widetilde\omega = 10^{-1}$, the displacement between the curves promoted by the perturbative theory and the Newtonian one is minimal. According to Table \eqref{tab:europa} with $\widetilde{\omega}=-0.1667$ the percentual difference is about $0.0063 \%$.

\subsection{Saturn-Prometheus System}
In the Saturn-Prometheus system, the average radius of the satellite will be $43.1$km, established as the midpoint between the values of $38$km and $48$km, which were selected as the lower and upper bounds, respectively.

\begin{table}[htb]\label{table3}
\caption{Percentage differences in the Roche limit for different values of $\widetilde{\omega}$. This results quantifies the curves shown in Fig. \ref{fig:prometheus} for the Saturn-Prometheus system}
    \label{tab:prometheus}
    \centering
    \begin{tabular}{c|c}
        $\widetilde\omega$ & $\text{Deviation from the Newtonian Roche limit}$ \\
        \hline
            $- 0.1667$ & $\approx 0.0016 \%$\\
            
            $- 0.01667$ & $\approx 0.015 \%$\\
            
            $- 0.001667$ & $\approx 0.15 \%$\\
        \end{tabular}
\end{table}

\begin{figure}[h]
\centering
\includegraphics[width=\linewidth]{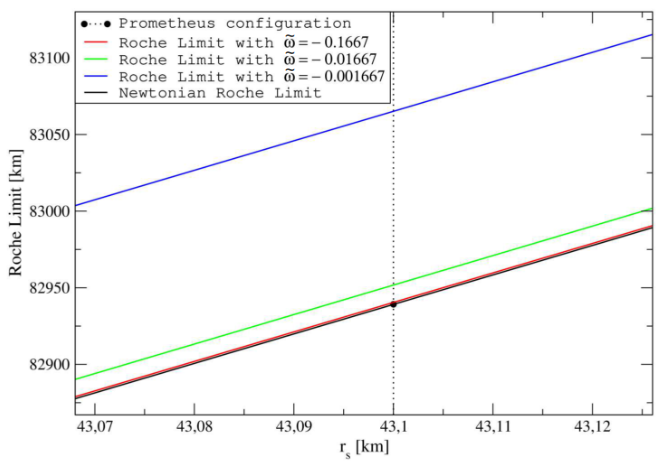}
\caption{Roche Limit Graph for different satellite radius values relative to a constant mass $m_s$. The different curves express the different values of $\widetilde\omega$. The dashed vertical line seeks to indicate the different Roche limits for $\widetilde\omega$ considering the radius of the Prometheus.}
\label{fig:prometheus}
\end{figure}

The Saturn-Prometheus system is composed of Saturn, which has a mass of $5.683 \times 10^{26}$ kg, and its natural satellite, Prometheus, with a mass of $1.595 \times 10^{17}$ kg. Once again, any value greater than $\widetilde\omega \approx -0.1667$ would result in a deviation from the expected results by the classical Roche Limit.

\section{Conclusions}

We have revisited the Newtonian-like theory inspired by the Brans-Dicke gravitational Lagrangian, as proposed in Reference \cite{jjtv}. In this study, our main focus relied on deriving the effective force experienced by massive test particles \eqref{force-pert} having as a free parameter $\omega$ or, equivalently, its redefinition $\widetilde{\omega}$ \eqref{widetilde}. The effective gravitational force resembles the exact form of the famous Manev potential. Subsequently, we applied this corrected gravitational force to conduct the classical test of Mercury's perihelion advance. The value $\widetilde{\omega}=-1/6$ is the one able to solve this test. Other bounds on $\widetilde{\omega}$ based on similar pericenter advance tests have also been imposed using different astrophysical systems.

The question now is how this value would impact other known results in the domain of celestial mechanics. In order to address this issue we investigated the impact of the value $\widetilde{\omega}=-1/6$ on the Roche limit. Our main finding is that this value has no significant impact, and the analysis demonstrates the same Newtonian behavior. In practise, the Roche limit is weakly sensitive to the $\omega$ value. This means that this theory can not damage local celestial dynamics. 

In brief, we have shown that the Newtonian theory with varying $G$, as proposed in Ref. \cite{jjtv}, remains robust when tested against the two constraints arising from celestial mechanics described above.

\acknowledgments{The authors thank FAPEMIG, FAPES and CNPq for financial support.}

\bibliography{Refs}

\begin{thebibliography}{21}%
\makeatletter
\providecommand \@ifxundefined [1]{%
 \@ifx{#1\undefined}
}%
\providecommand \@ifnum [1]{%
 \ifnum #1\expandafter \@firstoftwo
 \else \expandafter \@secondoftwo
 \fi
}%
\providecommand \@ifx [1]{%
 \ifx #1\expandafter \@firstoftwo
 \else \expandafter \@secondoftwo
 \fi
}%
\providecommand \natexlab [1]{#1}%
\providecommand \enquote  [1]{``#1''}%
\providecommand \bibnamefont  [1]{#1}%
\providecommand \bibfnamefont [1]{#1}%
\providecommand \citenamefont [1]{#1}%
\providecommand \href@noop [0]{\@secondoftwo}%
\providecommand \href [0]{\begingroup \@sanitize@url \@href}%
\providecommand \@href[1]{\@@startlink{#1}\@@href}%
\providecommand \@@href[1]{\endgroup#1\@@endlink}%
\providecommand \@sanitize@url [0]{\catcode `\\12\catcode `\$12\catcode
  `\&12\catcode `\#12\catcode `\^12\catcode `\_12\catcode `\%12\relax}%
\providecommand \@@startlink[1]{}%
\providecommand \@@endlink[0]{}%
\providecommand \url  [0]{\begingroup\@sanitize@url \@url }%
\providecommand \@url [1]{\endgroup\@href {#1}{\urlprefix }}%
\providecommand \urlprefix  [0]{URL }%
\providecommand \Eprint [0]{\href }%
\providecommand \doibase [0]{https://doi.org/}%
\providecommand \selectlanguage [0]{\@gobble}%
\providecommand \bibinfo  [0]{\@secondoftwo}%
\providecommand \bibfield  [0]{\@secondoftwo}%
\providecommand \translation [1]{[#1]}%
\providecommand \BibitemOpen [0]{}%
\providecommand \bibitemStop [0]{}%
\providecommand \bibitemNoStop [0]{.\EOS\space}%
\providecommand \EOS [0]{\spacefactor3000\relax}%
\providecommand \BibitemShut  [1]{\csname bibitem#1\endcsname}%
\let\auto@bib@innerbib\@empty
%</preamble>
\bibitem [{\citenamefont {Fabris}\ \emph
  {et~al.}(2021{\natexlab{a}})\citenamefont {Fabris}, \citenamefont {Gomes},
  \citenamefont {Toniato},\ and\ \citenamefont {Velten}}]{jjtv}%
  \BibitemOpen
  \bibfield  {author} {\bibinfo {author} {\bibfnamefont {J.~C.}\ \bibnamefont
  {Fabris}}, \bibinfo {author} {\bibfnamefont {T.}~\bibnamefont {Gomes}},
  \bibinfo {author} {\bibfnamefont {J.~D.}\ \bibnamefont {Toniato}},\ and\
  \bibinfo {author} {\bibfnamefont {H.}~\bibnamefont {Velten}},\ }\bibfield
  {title} {\bibinfo {title} {{Newtonian-like gravity with variable $G$}},\
  }\href {https://doi.org/10.1140/epjp/s13360-021-01146-z} {\bibfield
  {journal} {\bibinfo  {journal} {Eur. Phys. J. Plus}\ }\textbf {\bibinfo
  {volume} {136}},\ \bibinfo {pages} {143} (\bibinfo {year}
  {2021}{\natexlab{a}})},\ \Eprint {https://arxiv.org/abs/2009.04434}
  {arXiv:2009.04434 [gr-qc]} \BibitemShut {NoStop}%
\bibitem [{\citenamefont {Xue}\ \emph {et~al.}(2020)\citenamefont {Xue},
  \citenamefont {Liu}, \citenamefont {Li}, \citenamefont {Wu}, \citenamefont
  {Yang}, \citenamefont {Liu}, \citenamefont {Shao}, \citenamefont {Tu},
  \citenamefont {Hu},\ and\ \citenamefont {Luo}}]{10.1093/nsr/nwaa165}%
  \BibitemOpen
  \bibfield  {author} {\bibinfo {author} {\bibfnamefont {C.}~\bibnamefont
  {Xue}}, \bibinfo {author} {\bibfnamefont {J.-P.}\ \bibnamefont {Liu}},
  \bibinfo {author} {\bibfnamefont {Q.}~\bibnamefont {Li}}, \bibinfo {author}
  {\bibfnamefont {J.-F.}\ \bibnamefont {Wu}}, \bibinfo {author} {\bibfnamefont
  {S.-Q.}\ \bibnamefont {Yang}}, \bibinfo {author} {\bibfnamefont
  {Q.}~\bibnamefont {Liu}}, \bibinfo {author} {\bibfnamefont {C.-G.}\
  \bibnamefont {Shao}}, \bibinfo {author} {\bibfnamefont {L.-C.}\ \bibnamefont
  {Tu}}, \bibinfo {author} {\bibfnamefont {Z.-K.}\ \bibnamefont {Hu}},\ and\
  \bibinfo {author} {\bibfnamefont {J.}~\bibnamefont {Luo}},\ }\bibfield
  {title} {\bibinfo {title} {{Precision measurement of the Newtonian
  gravitational constant}},\ }\href {https://doi.org/10.1093/nsr/nwaa165}
  {\bibfield  {journal} {\bibinfo  {journal} {National Science Review}\
  }\textbf {\bibinfo {volume} {7}},\ \bibinfo {pages} {1803} (\bibinfo {year}
  {2020})},\ \Eprint
  {https://arxiv.org/abs/https://academic.oup.com/nsr/article-pdf/7/12/1803/38880653/nwaa165.pdf}
  {https://academic.oup.com/nsr/article-pdf/7/12/1803/38880653/nwaa165.pdf}
  \BibitemShut {NoStop}%
\bibitem [{\citenamefont {Marra}\ and\ \citenamefont
  {Perivolaropoulos}(2021)}]{PhysRevD.104.L021303}%
  \BibitemOpen
  \bibfield  {author} {\bibinfo {author} {\bibfnamefont {V.}~\bibnamefont
  {Marra}}\ and\ \bibinfo {author} {\bibfnamefont {L.}~\bibnamefont
  {Perivolaropoulos}},\ }\bibfield  {title} {\bibinfo {title} {Rapid transition
  of ${G}_{\mathrm{eff}}$ at ${z}_{t}\ensuremath{\simeq}0.01$ as a possible
  solution of the hubble and growth tensions},\ }\href
  {https://doi.org/10.1103/PhysRevD.104.L021303} {\bibfield  {journal}
  {\bibinfo  {journal} {Phys. Rev. D}\ }\textbf {\bibinfo {volume} {104}},\
  \bibinfo {pages} {L021303} (\bibinfo {year} {2021})}\BibitemShut {NoStop}%
\bibitem [{\citenamefont {Alestas}\ \emph {et~al.}(2022)\citenamefont
  {Alestas}, \citenamefont {Perivolaropoulos},\ and\ \citenamefont
  {Tanidis}}]{Alestas:2022xxm}%
  \BibitemOpen
  \bibfield  {author} {\bibinfo {author} {\bibfnamefont {G.}~\bibnamefont
  {Alestas}}, \bibinfo {author} {\bibfnamefont {L.}~\bibnamefont
  {Perivolaropoulos}},\ and\ \bibinfo {author} {\bibfnamefont {K.}~\bibnamefont
  {Tanidis}},\ }\bibfield  {title} {\bibinfo {title} {{Constraining a late time
  transition of G$_{eff}$ using low-z galaxy survey data}},\ }\href
  {https://doi.org/10.1103/PhysRevD.106.023526} {\bibfield  {journal} {\bibinfo
   {journal} {Phys. Rev. D}\ }\textbf {\bibinfo {volume} {106}},\ \bibinfo
  {pages} {023526} (\bibinfo {year} {2022})},\ \Eprint
  {https://arxiv.org/abs/2201.05846} {arXiv:2201.05846 [astro-ph.CO]}
  \BibitemShut {NoStop}%
\bibitem [{\citenamefont {Brans}\ and\ \citenamefont
  {Dicke}(1961)}]{Brans:1961sx}%
  \BibitemOpen
  \bibfield  {author} {\bibinfo {author} {\bibfnamefont {C.}~\bibnamefont
  {Brans}}\ and\ \bibinfo {author} {\bibfnamefont {R.}~\bibnamefont {Dicke}},\
  }\bibfield  {title} {\bibinfo {title} {{Mach's principle and a relativistic
  theory of gravitation}},\ }\href {https://doi.org/10.1103/PhysRev.124.925}
  {\bibfield  {journal} {\bibinfo  {journal} {Phys. Rev.}\ }\textbf {\bibinfo
  {volume} {124}},\ \bibinfo {pages} {925} (\bibinfo {year}
  {1961})}\BibitemShut {NoStop}%
\bibitem [{\citenamefont {Horndeski}(1974)}]{Horndeski:1974wa}%
  \BibitemOpen
  \bibfield  {author} {\bibinfo {author} {\bibfnamefont {G.~W.}\ \bibnamefont
  {Horndeski}},\ }\bibfield  {title} {\bibinfo {title} {{Second-order
  scalar-tensor field equations in a four-dimensional space}},\ }\href
  {https://doi.org/10.1007/BF01807638} {\bibfield  {journal} {\bibinfo
  {journal} {Int. J. Theor. Phys.}\ }\textbf {\bibinfo {volume} {10}},\
  \bibinfo {pages} {363} (\bibinfo {year} {1974})}\BibitemShut {NoStop}%
\bibitem [{\citenamefont {Duval}\ \emph {et~al.}(1991)\citenamefont {Duval},
  \citenamefont {Gibbons},\ and\ \citenamefont {Horvathy}}]{Duval:1990hj}%
  \BibitemOpen
  \bibfield  {author} {\bibinfo {author} {\bibfnamefont {C.}~\bibnamefont
  {Duval}}, \bibinfo {author} {\bibfnamefont {G.~W.}\ \bibnamefont {Gibbons}},\
  and\ \bibinfo {author} {\bibfnamefont {P.}~\bibnamefont {Horvathy}},\
  }\bibfield  {title} {\bibinfo {title} {{Celestial mechanics, conformal
  structures and gravitational waves}},\ }\href
  {https://doi.org/10.1103/PhysRevD.43.3907} {\bibfield  {journal} {\bibinfo
  {journal} {Phys. Rev. D}\ }\textbf {\bibinfo {volume} {43}},\ \bibinfo
  {pages} {3907} (\bibinfo {year} {1991})},\ \Eprint
  {https://arxiv.org/abs/hep-th/0512188} {arXiv:hep-th/0512188} \BibitemShut
  {NoStop}%
\bibitem [{\citenamefont {Dirac}(1937)}]{Dirac:1937ti}%
  \BibitemOpen
  \bibfield  {author} {\bibinfo {author} {\bibfnamefont {P.~A.}\ \bibnamefont
  {Dirac}},\ }\bibfield  {title} {\bibinfo {title} {{The Cosmological
  constants}},\ }\href {https://doi.org/10.1038/139323a0} {\bibfield  {journal}
  {\bibinfo  {journal} {Nature}\ }\textbf {\bibinfo {volume} {139}},\ \bibinfo
  {pages} {323} (\bibinfo {year} {1937})}\BibitemShut {NoStop}%
\bibitem [{\citenamefont {Dirac}(1938)}]{Dirac:1938mt}%
  \BibitemOpen
  \bibfield  {author} {\bibinfo {author} {\bibfnamefont {P.~A.}\ \bibnamefont
  {Dirac}},\ }\bibfield  {title} {\bibinfo {title} {{New basis for
  cosmology}},\ }\href {https://doi.org/10.1098/rspa.1938.0053} {\bibfield
  {journal} {\bibinfo  {journal} {Proc. Roy. Soc. Lond. A}\ }\textbf {\bibinfo
  {volume} {A165}},\ \bibinfo {pages} {199} (\bibinfo {year}
  {1938})}\BibitemShut {NoStop}%
\bibitem [{\citenamefont {Fabris}\ \emph
  {et~al.}(2021{\natexlab{b}})\citenamefont {Fabris}, \citenamefont {Ottoni},
  \citenamefont {Toniato},\ and\ \citenamefont {Velten}}]{Fabris:2021qkp}%
  \BibitemOpen
  \bibfield  {author} {\bibinfo {author} {\bibfnamefont {J.~C.}\ \bibnamefont
  {Fabris}}, \bibinfo {author} {\bibfnamefont {T.}~\bibnamefont {Ottoni}},
  \bibinfo {author} {\bibfnamefont {J.~D.}\ \bibnamefont {Toniato}},\ and\
  \bibinfo {author} {\bibfnamefont {H.}~\bibnamefont {Velten}},\ }\bibfield
  {title} {\bibinfo {title} {{Stellar Structure in a Newtonian Theory with
  Variable G}},\ }\href {https://doi.org/10.3390/physics3040071} {\bibfield
  {journal} {\bibinfo  {journal} {MDPI Physics}\ }\textbf {\bibinfo {volume}
  {3}},\ \bibinfo {pages} {1123} (\bibinfo {year} {2021}{\natexlab{b}})},\
  \Eprint {https://arxiv.org/abs/2109.08687} {arXiv:2109.08687 [gr-qc]}
  \BibitemShut {NoStop}%
\bibitem [{\citenamefont {Poisson}\ and\ \citenamefont
  {Will}(2014)}]{poisson_will_2014}%
  \BibitemOpen
  \bibfield  {author} {\bibinfo {author} {\bibfnamefont {E.}~\bibnamefont
  {Poisson}}\ and\ \bibinfo {author} {\bibfnamefont {C.~M.}\ \bibnamefont
  {Will}},\ }\href {https://doi.org/10.1017/CBO9781139507486} {\emph {\bibinfo
  {title} {Gravity: Newtonian, Post-Newtonian, Relativistic}}}\ (\bibinfo
  {publisher} {Cambridge University Press},\ \bibinfo {year}
  {2014})\BibitemShut {NoStop}%
\bibitem [{\citenamefont {Park}\ \emph {et~al.}(2017)\citenamefont {Park},
  \citenamefont {Folkner}, \citenamefont {Konopliv}, \citenamefont {Williams},
  \citenamefont {Smith},\ and\ \citenamefont {Zuber}}]{Park_2017}%
  \BibitemOpen
  \bibfield  {author} {\bibinfo {author} {\bibfnamefont {R.~S.}\ \bibnamefont
  {Park}}, \bibinfo {author} {\bibfnamefont {W.~M.}\ \bibnamefont {Folkner}},
  \bibinfo {author} {\bibfnamefont {A.~S.}\ \bibnamefont {Konopliv}}, \bibinfo
  {author} {\bibfnamefont {J.~G.}\ \bibnamefont {Williams}}, \bibinfo {author}
  {\bibfnamefont {D.~E.}\ \bibnamefont {Smith}},\ and\ \bibinfo {author}
  {\bibfnamefont {M.~T.}\ \bibnamefont {Zuber}},\ }\bibfield  {title} {\bibinfo
  {title} {Precession of mercury's perihelion from ranging to the {MESSENGER
  Spacecraft}},\ }\href {https://doi.org/10.3847/1538-3881/aa5be2} {\bibfield
  {journal} {\bibinfo  {journal} {The Astronomical Journal}\ }\textbf {\bibinfo
  {volume} {153}},\ \bibinfo {pages} {121} (\bibinfo {year}
  {2017})}\BibitemShut {NoStop}%
\bibitem [{\citenamefont {Krasinsky}\ \emph {et~al.}(1993)\citenamefont
  {Krasinsky}, \citenamefont {Pitjeva}, \citenamefont {Sveshnikov},\ and\
  \citenamefont {Chunayeva}}]{Krasinsky:1993}%
  \BibitemOpen
  \bibfield  {author} {\bibinfo {author} {\bibfnamefont {G.~A.}\ \bibnamefont
  {Krasinsky}}, \bibinfo {author} {\bibfnamefont {E.~V.}\ \bibnamefont
  {Pitjeva}}, \bibinfo {author} {\bibfnamefont {M.~L.}\ \bibnamefont
  {Sveshnikov}},\ and\ \bibinfo {author} {\bibfnamefont {L.~I.}\ \bibnamefont
  {Chunayeva}},\ }\bibfield  {title} {\bibinfo {title} {The motion of major
  planets from observations 1769--1988 and some astronomical constants},\
  }\href {https://doi.org/10.1007/BF00694392} {\bibfield  {journal} {\bibinfo
  {journal} {Celestial Mechanics and Dynamical Astronomy}\ }\textbf {\bibinfo
  {volume} {55}},\ \bibinfo {pages} {1} (\bibinfo {year} {1993})}\BibitemShut
  {NoStop}%
\bibitem [{\citenamefont {Pitjeva}(2001)}]{Pitjeva:2001}%
  \BibitemOpen
  \bibfield  {author} {\bibinfo {author} {\bibfnamefont {E.~V.}\ \bibnamefont
  {Pitjeva}},\ }\bibfield  {title} {\bibinfo {title} {Modern numerical
  ephemerides of planets and the importance of ranging observations for their
  creation},\ }\href {https://doi.org/10.1023/A:1012289530641} {\bibfield
  {journal} {\bibinfo  {journal} {Celestial Mechanics and Dynamical Astronomy}\
  }\textbf {\bibinfo {volume} {80}},\ \bibinfo {pages} {249} (\bibinfo {year}
  {2001})}\BibitemShut {NoStop}%
\bibitem [{\citenamefont {Lucchesi}\ and\ \citenamefont
  {Peron}(2010)}]{Lucchesi:2010}%
  \BibitemOpen
  \bibfield  {author} {\bibinfo {author} {\bibfnamefont {D.~M.}\ \bibnamefont
  {Lucchesi}}\ and\ \bibinfo {author} {\bibfnamefont {R.}~\bibnamefont
  {Peron}},\ }\bibfield  {title} {\bibinfo {title} {Accurate measurement in the
  field of the earth of the general-relativistic precession of the lageos ii
  pericenter and new constraints on non-newtonian gravity},\ }\bibfield
  {journal} {\bibinfo  {journal} {Physical Review Letters}\ }\textbf {\bibinfo
  {volume} {105}},\ \href {https://doi.org/10.1103/physrevlett.105.231103}
  {10.1103/physrevlett.105.231103} (\bibinfo {year} {2010})\BibitemShut
  {NoStop}%
\bibitem [{\citenamefont {Abuter}\ \emph {et~al.}(2020)\citenamefont {Abuter},
  \citenamefont {Amorim}, \citenamefont {Bauböck}, \citenamefont {Berger},
  \citenamefont {Bonnet}, \citenamefont {Brandner}, \citenamefont {Cardoso},
  \citenamefont {Clénet}, \citenamefont {de~Zeeuw},\ and\ \citenamefont
  {et~al.}}]{Abuter:2020}%
  \BibitemOpen
  \bibfield  {author} {\bibinfo {author} {\bibfnamefont {R.}~\bibnamefont
  {Abuter}}, \bibinfo {author} {\bibfnamefont {A.}~\bibnamefont {Amorim}},
  \bibinfo {author} {\bibfnamefont {M.}~\bibnamefont {Bauböck}}, \bibinfo
  {author} {\bibfnamefont {J.~P.}\ \bibnamefont {Berger}}, \bibinfo {author}
  {\bibfnamefont {H.}~\bibnamefont {Bonnet}}, \bibinfo {author} {\bibfnamefont
  {W.}~\bibnamefont {Brandner}}, \bibinfo {author} {\bibfnamefont
  {V.}~\bibnamefont {Cardoso}}, \bibinfo {author} {\bibfnamefont
  {Y.}~\bibnamefont {Clénet}}, \bibinfo {author} {\bibfnamefont {P.~T.}\
  \bibnamefont {de~Zeeuw}},\ and\ \bibinfo {author} {\bibnamefont {et~al.}},\
  }\bibfield  {title} {\bibinfo {title} {Detection of the schwarzschild
  precession in the orbit of the star s2 near the galactic centre massive black
  hole},\ }\href {https://doi.org/10.1051/0004-6361/202037813} {\bibfield
  {journal} {\bibinfo  {journal} {Astronomy \& Astrophysics}\ }\textbf
  {\bibinfo {volume} {636}},\ \bibinfo {pages} {L5} (\bibinfo {year}
  {2020})}\BibitemShut {NoStop}%
\bibitem [{\citenamefont {Famaey}\ and\ \citenamefont
  {McGaugh}(2012)}]{Famaey2012}%
  \BibitemOpen
  \bibfield  {author} {\bibinfo {author} {\bibfnamefont {B.}~\bibnamefont
  {Famaey}}\ and\ \bibinfo {author} {\bibfnamefont {S.~S.}\ \bibnamefont
  {McGaugh}},\ }\bibfield  {title} {\bibinfo {title} {Modified {{Newtonian
  Dynamics}} ({{MOND}}): {{Observational Phenomenology}} and {{Relativistic
  Extensions}}},\ }\href {https://doi.org/10.12942/lrr-2012-10} {\bibfield
  {journal} {\bibinfo  {journal} {Living Reviews in Relativity}\ }\textbf
  {\bibinfo {volume} {15}},\ \bibinfo {pages} {10} (\bibinfo {year}
  {2012})}\BibitemShut {NoStop}%
\bibitem [{\citenamefont {Milgrom}(2015)}]{Milgrom2015}%
  \BibitemOpen
  \bibfield  {author} {\bibinfo {author} {\bibfnamefont {M.}~\bibnamefont
  {Milgrom}},\ }\bibfield  {title} {\bibinfo {title} {Mond theory},\ }\href
  {https://doi.org/10.1139/cjp-2014-0211} {\bibfield  {journal} {\bibinfo
  {journal} {Canadian Journal of Physics}\ }\textbf {\bibinfo {volume} {93}},\
  \bibinfo {pages} {107} (\bibinfo {year} {2015})}\BibitemShut {NoStop}%
\bibitem [{\citenamefont {{Bekenstein}}\ and\ \citenamefont
  {{Milgrom}}(1984)}]{Bekenstein1984}%
  \BibitemOpen
  \bibfield  {author} {\bibinfo {author} {\bibfnamefont {J.}~\bibnamefont
  {{Bekenstein}}}\ and\ \bibinfo {author} {\bibfnamefont {M.}~\bibnamefont
  {{Milgrom}}},\ }\bibfield  {title} {\bibinfo {title} {{Does the missing mass
  problem signal the breakdown of Newtonian gravity?}},\ }\href
  {https://doi.org/10.1086/162570} {\bibfield  {journal} {\bibinfo  {journal}
  {\apj}\ }\textbf {\bibinfo {volume} {286}},\ \bibinfo {pages} {7} (\bibinfo
  {year} {1984})}\BibitemShut {NoStop}%
\bibitem [{\citenamefont {Touboul}\ \emph {et~al.}(2022)\citenamefont {Touboul}
  \emph {et~al.}}]{microscope}%
  \BibitemOpen
  \bibfield  {author} {\bibinfo {author} {\bibfnamefont {P.}~\bibnamefont
  {Touboul}} \emph {et~al.} (\bibinfo {collaboration} {MICROSCOPE
  Collaboration}),\ }\bibfield  {title} {\bibinfo {title} {{$MICROSCOPE$
  Mission}: Final results of the test of the equivalence principle},\ }\href
  {https://doi.org/10.1103/PhysRevLett.129.121102} {\bibfield  {journal}
  {\bibinfo  {journal} {Phys. Rev. Lett.}\ }\textbf {\bibinfo {volume} {129}},\
  \bibinfo {pages} {121102} (\bibinfo {year} {2022})}\BibitemShut {NoStop}%
\bibitem [{\citenamefont {Hofmann}\ and\ \citenamefont
  {Müller}(2018)}]{Hofmann_2018}%
  \BibitemOpen
  \bibfield  {author} {\bibinfo {author} {\bibfnamefont {F.}~\bibnamefont
  {Hofmann}}\ and\ \bibinfo {author} {\bibfnamefont {J.}~\bibnamefont
  {Müller}},\ }\bibfield  {title} {\bibinfo {title} {Relativistic tests with
  lunar laser ranging},\ }\href {https://doi.org/10.1088/1361-6382/aa8f7a}
  {\bibfield  {journal} {\bibinfo  {journal} {Classical and Quantum Gravity}\
  }\textbf {\bibinfo {volume} {35}},\ \bibinfo {pages} {035015} (\bibinfo
  {year} {2018})}\BibitemShut {NoStop}%
\end{thebibliography}%

\end{document}